\documentclass[floatfix,showkeys,twocolumn,citeautoscript,showpacs,amsmath,amssymb,epsf,aps,prb]{revtex4}

\usepackage{graphics}
\usepackage{epsfig}
\usepackage{longtable}
\usepackage{dcolumn}
\usepackage{bm}
\usepackage{amssymb}

\def\etal{{\em et al. }}
\def\trip_b{$^3$B$_1$ }
\def\sing_a{$^1$A$_1$ }
\def\t_b{$^3$B$_1$ }
\def\s_a{$^1$A$_1$ }
\def\af{$\alpha$ }

\begin{document}

\title{ On optimal values of \af for the analytic Hartree-Fock-Slater method}

\author{Rajendra R. Zope}
\email{rzope@alchemy.nrl.navy.mil}
\affiliation{Department of Chemistry, George Washington University, Washington DC, 20052}

\author{Brett  I. Dunlap}
\email{dunlap@nrl.navy.mil}
\affiliation{Code 6189, Theoretical Chemistry Section, US Naval Research Laboratory
Washington, DC 20375}


\date{\today}

\begin{abstract}
     We have examined the performance of the analytic Hartree-Fock-Slater (HFS)  method
for various \af values and empiricaly determined the optimal \af value by minimizing
the mean absolute error (MAE) in atomization energies of the G2 set of molecules.
At the optimal $\alpha$  the HFS method's performance is far superior with the  MAE of
14 kcal/mol than that of the local density approximation (MAE $\sim$ 36 kcal/mol)
or the Hartree-Fock theory (MAE $\sim$ 78 kcal/mol).   The HFS exchange
functional with $\alpha = 0.7091 $ performs significantly better than the
Kohn-Sham exchange functional for equally weighted atoms H-Kr.
We speculate that use of this single \af value may
be useful in parametrization of empirical exchange-correlation functionals.
\end{abstract}

\keywords{ analytic density functional theory, exchange 
potential, Slater's X$\alpha$, Hartree-Fock-Slater}

\maketitle

The problems with numerical integration in quantum mechanical calculations are well
known\cite{Cook1,JF93}.  Thus, all {\it ab initio} electronic structure calculations 
can be divided into two classes.  In the first class, historically called {\it ab initio},
all quantum mechanical matrix elements are computed to machine precision\cite{Pople98}. 
The second class
of electronic structure calculations require numerical integration and machine-precision
matrix elements are totally impractical, and thus, except for atoms, machine-precision
energies are out of the question.  Until very recently  this second class included almost all
density-functional calculations\cite{Kohn99}.  
Recently, fully analytic Hartree-Fock-Slater\cite{Slater51} (HFS) variant of the 
density functional theory (DFT) was implemented using Gaussian basis sets\cite{Cook1}.  
This approach employs fitting of the potential to 
integrable functional form, rather 
than by fitting or integration on numerical grid. The technique is  computationally very
efficient in comparison 
with the grid-based implementation and provides smooth potential energy surfaces and 
exact energy gradients\cite{Cook1}.
We have recently  extended this scheme to allow for the atom-dependent exchange 
parameters \af that scale the exchange potential by means of a muffin-tin (MT)-like 
approach\cite{Dunlap03}. In our method matrix elements are computed to machine 
accuracy. Further, in contrast to earlier MT implementation,
here the energy is both meaningful and stationary. One can require that 
atoms dissociate into their exact experimental rather than  approximate Hartree-Fock 
(HF) electronic energies. This approach\cite{ZB_BN} when applied to a standard 
set of molecules that 
are used in performance tests of DFT models yields  results that are intermediate 
between either the local density approximation (LDA) or the HF appoxumation and 
more sophisticated hybrid or generalized gradient approximations (GGA)\cite{ZB_SR}.

In the HFS\cite{Slater51} model, the {\sl nonlocal} exchange potential in the
Hartree-Fock method is replaced by a {\it local} exchange potential that is given by
\begin{equation}
   v_{x} (\vec{r}) = - \frac{3}{2}  \Bigl( \frac{3}{\pi} \Bigr )^{1/3}
  \alpha   \rho^{1/3}.
   \label{Eq:Exc}
\end{equation}
Here, the parameter $\alpha$, called Slater's statistical
exchange parameter, is unity.  Similar  expression for the exchange
energy of the homogeneous electron gas was obtained earlier by Dirac\cite{Dirac30}.
Later, G\'asp\'ar \cite{Gaspar} and Kohn-Sham \cite{KS65}  (GKS) obtained the value of $2/3$
for \af by variationally minimizing the total energy functional.
In the following years, \af was taken purely as an adjustable
parameter to obtain desired atomic properties.\cite{Schwarz,SS78,Connolly,Fliszar1}
The first\cite{BR73}  HFS calculations with meaningful numerically integrated total energies used
a uniform \af value of 0.7.
Since then the HFS method has come to mean this \af value.
Later, the electronic structure calculations
using the LDA by  showed that
the LDA give similar\cite{BR78} but not superior \cite{GHJ77} binding energies to the
HFS method. Several studies since then have shown that the LDA has a general tendency
to overbind\cite{DFT:Koch}.

    HF theory being analytic allows  cheap geometry optimization despite its $N^4$ 
cost. In an analytic method one optimizes tens of linear-combination-of-atomic-orbital parameters
per atom, rather than hundreds of plane-waves per pseudoatom, or thousands of numerical integration 
points per all-electron atom. With or without the MT-like advance,  an $N^3$ analytic 
method might prove to be a practical geometry-optimization tool if appropriate choice(s)
of the exchange parameter(s) is(are) made. 
  In this article we asses the performance of analytic HF model for the 
GKS and the Slater values of \af by computing the mean absolute error (MAE) in atomization 
energies of a set of 56 molecules (G2 set). We then determine the optimal 
value of \af by minimizing the MAE for the G2 set of molecules. The calculations are 
performed for various basis sets in order to study the basis set dependence of the 
optimal \af value. 
Our calculations show that the analytic HFS model with the optimal 
\af value performs better than the HF theory or the LDA and hence  provides 
a computationally efficient scheme to study large systems at modest accuracy. Furthermore,
by minimizing the MAE between the HF and the HFS  total energies for atoms H through Kr,
we find that best performance of the exchange functional in Eq.(\ref{Eq:Exc})
is obtained for $\alpha = 0.7091 $.

       Our calculations in  the Slater-Roothaan (SR) method  require using the 
Gaussian basis sets to fit the orbitals and the Kohn-Sham potential. We 
have used the valence triple-$\zeta$ (TZ) 6-311G**\cite{O1,O2} and the DGauss\cite{AW91} 
valence double-$\zeta$\cite{GSAW92} basis set (DZVP) for the orbitals.  
The {\sl s-}type fitting bases are obtained by scaling the {\sl s-} part of
the orbital basis \cite{Dunlap79}. For the non-zero angular momentum components
the resolution-of-the-identity-J (RI-J)\cite{EWTR97} and A2 \cite{AW91} basis sets are
used for the Kohn-Sham potential fitting.  Thus, four sets  6311G**/RI-J, 6311G**/A2,
DZVP/RI-J, and  DZVP/A2 of bases were used for optimizing the  \af value.  The 
molecules were optimized using the Broyden-Fletcher-Goldfarb-Shanno (BFGS)
algorithm\cite{BFGS}. The \af minimization was performed using powerful Perl scripts
that drive the analytic DFT code.

    Our first attempt to determine the optimal \af is based on the atomic calculation.
These calculations are numerical and therefore are free from the basis sets effects.  Here
we minimize the MAE in the HF and the HFS total energies for atoms H through Ar.
The minimum occurs for $\alpha =  0.7267 $ with the MAE of 0.101 a.u. The optimal
\af value decreases slightly to 0.7091 when the target set is extended to include 
the atoms up to krypton. At this value the MAE is 0.33 au. These errors are an order 
of magnitude smaller than the MAE (2.38 a.u.) for the  $\alpha = 2/3$.
The exchange functional in Eq. (\ref{Eq:Exc}) with $\alpha = 0.7091$ is therefore 
better approximation than the GKS functional, at least for the atomic systems.

\begin{table*}     
\begin{ruledtabular}    
\caption{The atomization energies D$_0$ (kcal/mol) for the 56 set of molecules   
for two different basis sets.  The two  basis sets chosen are I: 6311G**/RIJ,  II: DZVP/A2.   
The ${\alpha}$  values are Gaspar-Kohn-Sham's alpha (=0.66666667),  Slater's alpha (=1.0000),    
and the optimal ${\alpha}$ for which mean absolute error is minimum.  The last    
column contains the {\em exact} values. }   
\label{tab:m56}       
\begin{tabular} {lrrrrrrr}   
                &  Basis  I    &  Basis I &  Basis I & Basis II & Basis II & Basis II &         \\   
$\alpha$        &  0.666667    &  1.00000 &  0.70650  &  0.666667 & 1.00000  & 0.69800  &   Exact \\   
 \hline           \\ 
 H$_2$      	&    81.56  	&   91.53  &   82.76  &    84.16 &   94.13  &   84.90  &  103.50    \\ 
 LiH      	&    33.44  	&   48.59  &   34.86  &    30.89 &   41.90  &   31.49  &   56.00    \\ 
 BeH      	&    46.62  	&   79.40  &   50.09  &    25.52 &   46.23  &   27.11  &   46.90    \\ 
 CH      	&    63.16  	&   74.57  &   64.46  &    64.95 &   75.70  &   65.78  &   79.90    \\ 
 CH$_2$ (\t_b)  & 172.67  	&  242.62  &  180.69  &   175.69 &  244.94  &  181.41  &   179.60    \\ 
 CH$_2$ (\s_a)  & 144.13  	&  187.45  &  149.03  &   148.50 &  190.34  &  151.92  &   170.60   \\ 
 CH$_3$      	&   269.96  	&  359.08  &  280.32  &   274.91 &  362.50  &  282.28  &  289.20    \\ 
 CH$_4$      	&   369.54  	&  484.63  &  382.87  &   375.72 &  490.08  &  385.38  &  392.50    \\ 
 NH      	&    62.39  	&   77.94  &   64.13  &    65.12 &   79.59  &   66.21  &   79.00    \\ 
 NH$_2$      	&   144.46  	&  186.28  &  149.20  &   150.51 &  189.84  &  153.69  &  170.00    \\ 
 NH$_3$      	&   247.52  	&  327.10  &  256.68  &   256.44 &  332.45  &  262.79  &  276.70    \\ 
 OH      	&    91.11  	&  116.00  &   94.03  &    94.17 &  116.74  &   96.03  &  101.30    \\ 
 H$_2$O      	&   205.36  	&  281.07  &  214.13  &   213.20 &  283.90  &  219.17  &  219.30    \\ 
 HF      	&   129.94  	&  181.58  &  136.01  &   136.08 &  183.42  &  140.19  &  135.20    \\ 
 Li$_2$      	&     6.75  	&    7.78  &    6.65  &     5.74 &    9.69  &    5.66  &   24.00    \\ 
 LiF      	&   127.73  	&  200.55  &  135.59  &   118.91 &  192.70  &  124.75  &  137.76    \\ 
 C$_2$H$_2$    	&   380.64  	&  508.89  &  395.62  &   375.39 &  495.61  &  385.98  &  388.90    \\ 
 C$_2$H$_4$    	&   514.63  	&  688.39  &  534.80  &   519.43 &  688.06  &  534.03  &  531.90    \\ 
 C$_2$H$_6$    	&   640.65  	&  858.65  &  665.79  &   649.78 &  864.88  &  668.10  &  666.30    \\ 
 CN      	&   175.85  	&  226.64  &  180.87  &   168.49 &  210.50  &  171.89  &  176.60    \\ 
 HCN      	&   291.80  	&  371.10  &  300.99  &   284.76 &  354.24  &  291.08  &  301.80    \\ 
 CO      	&   262.44  	&  327.45  &  269.96  &   251.66 &  307.16  &  257.01  &  256.20    \\ 
 HCO      	&   279.07  	&  370.60  &  289.35  &   274.94 &  359.47  &  282.37  &  270.30    \\ 
 H$_2$CO      	&   357.03  	&  476.46  &  370.63  &   356.06 &  469.07  &  365.93  &  357.20    \\ 
 CH$_3$OH      	&   468.42  	&  640.15  &  487.97  &   476.56 &  643.49  &  490.71  &  480.80    \\ 
 N$_2$      	&   206.16  	&  233.59  &  209.47  &   196.84 &  212.00  &  198.89  &  225.10    \\ 
 N$_2$H$_4$    	&   371.99  	&  512.35  &  387.70  &   386.51 &  519.20  &  397.56  &  405.40    \\ 
 NO      	&   153.63  	&  186.81  &  157.31  &   146.96 &  171.38  &  149.52  &  150.10    \\ 
 O$_2$      	&   144.97  	&  191.03  &  149.76  &   141.18 &  183.86  &  145.09  &  118.00    \\ 
 H$_2$O$_2$    	&   255.11  	&  356.29  &  266.06  &   262.57 &  361.22  &  270.64  &  252.30    \\ 
 F$_2$      	&    60.99  	&   86.35  &   63.10  &    60.70 &   91.54  &   63.08  &   36.90    \\ 
 CO$_2$      	&   413.28  	&  554.85  &  429.25  &   398.39 &  524.83  &  409.85  &  381.90    \\ 
 SiH$_2$ (\s_a) &  116.69  	&  140.69  &  119.57  &   119.85 &  146.07  &  121.96  &  144.40    \\ 
 SiH$_2$ (\t_b) & 110.16  	&  149.90  &  114.64  &   112.65 &  153.60  &  115.87  &  123.40    \\ 
 SiH$_3$      	&   179.61  	&  229.59  &  185.47  &   183.25 &  234.73  &  187.39  &  214.00    \\ 
 SiH$_4$      	&   260.34  	&  326.68  &  268.22  &   264.75 &  333.84  &  270.39  &  302.80    \\ 
 PH$_2$      	&   119.12  	&  139.95  &  121.62  &   123.02 &  147.19  &  124.89  &  144.70    \\ 
 PH$_3$      	&   191.36  	&  229.18  &  195.93  &   197.19 &  240.50  &  200.61  &  227.40    \\ 
 H$_2$S      	&   154.22  	&  188.60  &  158.33  &   158.87 &  199.34  &  162.23  &  173.20    \\ 
 HCl      	&    95.69  	&  118.24  &   98.50  &    96.37 &  123.43  &   98.65  &  102.20    \\ 
 Na$_2$      	&     5.47  	&    3.85  &    5.20  &     5.59 &    4.66  &    5.42  &   16.60    \\ 
 Si$_2$      	&    68.36  	&   90.14  &   70.80  &    68.42 &   92.93  &   70.48  &   74.00    \\ 
 P$_2$      	&    91.33  	&  101.21  &   92.69  &    91.67 &  103.08  &   92.73  &  116.10    \\ 
 S$_2$      	&   102.03  	&  137.80  &  106.19  &   103.03 &  141.76  &  106.46  &  100.70    \\ 
 Cl$_2$      	&    57.18  	&   92.08  &   60.88  &    58.37 &   95.81  &   61.29  &   57.20    \\ 
 NaCl      	&    80.52  	&  121.13  &   84.89  &    81.00 &  127.66  &   84.84  &   97.50    \\ 
 SiO      	&   182.85  	&  239.02  &  189.13  &   182.19 &  239.66  &  187.07  &  190.50    \\ 
 CS      	&   165.02  	&  202.06  &  169.50  &   163.97 &  202.37  &  167.64  &  169.50    \\ 
 SO      	&   128.41  	&  177.03  &  133.65  &   133.04 &  183.35  &  137.30  &  123.50    \\ 
 ClO      	&    69.56  	&  104.25  &   72.98  &    74.70 &  111.20  &   77.59  &   63.30    \\ 
 ClF      	&    69.97  	&  109.07  &   73.98  &    74.16 &  116.73  &   77.61  &   60.30    \\ 
 Si$_2$H$_6$   	&   438.03  	&  562.19  &  452.72  &   445.26 &  576.48  &  456.04  &  500.10    \\ 
 CH$_3$Cl      	&   361.67  	&  484.96  &  375.92  &   366.70 &  492.51  &  377.46  &  371.00    \\ 
 H$_3$CSH      	&   423.74  	&  560.88  &  439.59  &   432.27 &  574.77  &  444.35  &  445.10    \\ 
 HOCl      	&   156.11  	&  222.91  &  163.21  &   163.22 &  234.11  &  168.94  &  156.30    \\ 
 SO$_2$      	&   250.06  	&  356.02  &  261.65  &   254.36 &  360.56  &  263.42  &  254.00    \\ 
\hline  
 mean absolute  &    15.9   	&   55.2   & 13.5     &   14.5   &  55.3    &   12.8   &            \\ 
 \end{tabular}         
\end{ruledtabular}    
\end{table*}

  We now examine the performance of the HFS model for the GKS and Slater \af values.
  The performance of the analytic HFS model in prediction of the atomization energies 
for the G2 set of molecules is given in Table \ref{tab:m56}. The computation of atomization
energies is a stringent test for computational models and has been routinely used in the appraisal
of the computational models. The G2 set of molecules used in the performance analysis is 
the set of 54 molecules in 56 electronic states due to Pople and coworkers \cite{Pople1}
 and 
is often used for performance tests\cite{Becke93}.  It is apparent from the Table \ref{tab:m56}
that the
errors are considerably smaller for the GKS value of \af than for Slater ${\alpha}$.
The atomization
energies are overestimated for Slater \af with a mean error of 52.4 kcal/mol while 
for the GKS \af the molecules are by and large underbound. The minimization of the MAE 
leads to the  \af value close to 0.7. The atomization energies for the G2 set at the 
optimal values are also shown in Table  \ref{tab:m56}. Going from the GKS \af to the optimal \af,
the MAE reduces by  about 2 kcal/mol while the mean errors decreases by  6-7 kcal/mol.
The reduction in the mean error mainly occurs because it changes sign for more molecules.
For the GKS $\alpha$, the error is maximum (-61.49 kcal/mol) for the Si$_2$H$_6$. It is also
maximal  for Si$_2$H$_6$ at the optimal \af.   CO$_2$ is 
another molecule for which the error is comparable to this error at the optimal \af.

    In order to investigate the role of basis sets on the optimal $\alpha$ value,
we optimized the \af for four different basis sets. The optimal \af 
values are 0.70650, 0.69937, 0.7032, and  0.698 for the 6311G**/RI-J, 6311G**/A2, DZVP/RI-J, 
and DZVP/A2, respectively. The MAE (in units of kcal/mol) 
at these optimal values are 13.5, 13.4, 12.8, and 12.8,
respectively. All optimal values are close to 0.7 and are effectively insensitive to  
the orbital basis sets. A small dependence on the fitting basis set is however noticeable.
The best performance is obtained for the DZVP/A2 basis set for which the MAE is 12.8 kcal/mol.

     We also carried out the performance test of the model in predicting the bond distances. 
For this purpose we selected all (15 in total) diatomic molecules belonging to the G2 set. 
Our results show that
the MAE in bond distances also is smaller at the optimal \af value. For the  6311G**/RI-J the 
MAE at optimal value is 0.019 \AA,   0.013 ~\AA \, smaller than at the GKS \af (0.032 \AA).
The basis set effects show that the larger  6311G**/RI-J performs better than the DZVP/A2 
basis. The MAE at their optimal \af values for these two bases are  0.019 and 0.048 \AA, 
respectively. These are comparable or better than the LDA ( 0.024 \AA)  or HF (0.028 \AA) 
errors\cite{Pople_DFT}.

   In Table \ref{table:err} we have summarized the results of present calculations. In 
comparison with the  LDA  or HF the analytic HFS model performance is significantly 
better.  Its performance is even better than the SR-HF  or the SR-Exact-Atomic models.
These models are similar to the present one but make use of atom dependent  ${\alpha},$
which in case of SR-HF and  SR-Exact-Atomic uses \af values that give the HF 
atomic and the exact energies for atomic systems, respectively. The overall 
improvement in the performance obtained here by minimizing the MAE in atomization
energies also suggest that the SR model can also be similarly improved by multidimensional
minimization of MAE in the \af space.  
There are several 
density functional computational schemes that use the generalized gradient 
approximation (GGA), the hybrid GGA or meta GGA(See for example, Ref. ~\onlinecite{DFT:Koch}).
%
%
The accuracies  of these models for the G2 set  range from 3-8 kcal/mol,
but to date they require numerical treatment.
%
%
%
%
Although its analytic implementation is 
computationally most efficient, the optimal values can also be used in 
any existing density functional code, albeit with some reduction in computational
performance. 
It should also be borne in mind that the G2 set used in obtaining optimal 
\af value contains small molecules consisting of atoms belonging to the first 
and second rows of the periodic table.  

We have also examined the performance of the analytic HFS model for the extended
G2 set containing 148 molecules.
Reoptimizing the \af in order to minimize the MAE for this larger G2 set moves
optimal \af significantly far in the direction of the GKS's \af value. The 
analysis of errors 
for individual molecules in this dataset shows that this occurs due to the 
presence of a large percent of  molecules containing fluorine in the extended G2 dataset. 
The errors for these molecules are lowered by decreasing the \af value below 0.7. 
This is consistent with our earlier finding that the exact atomization 
of fluorine dimer is obtained for much smaller \af value of 0.3\cite{ZB_SR}.
This again brings out the limitation of the uniform \af  HFS method and shows
that the analytic SR method has a scope for improvement. 
It appears from the minimization of errors of the G2 and the extended G2 data 
sets and error analysis,  as well as from the minimization of the total 
total atomic energies that overall the value close to 0.7 is probably 
the right choice for the optimal \af in the uniform \af calculations.



\begin{table}
\caption{ The mean absolute error (MAE) (kcal/mol) in the atomization energy of 56 molecules 
belonging to the G2 set is compared within different models.  The numbers for the
 SR-HF and SR-{\em Exact-Atomic} are for the Slater-Roothaan model with 
Hartree-Fock \af values and the \af values that give the {\em exact-atomic}
(See text for more details). The results of the more complex PBE GGA functional are
also  included for comparison.}
\label{table:err}
\begin{tabular}{lll}
\hline
   Model &    Basis        & MAE  \\
\hline
   Hartree-Fock theory   &                 & 78  ~Ref.\onlinecite{Becke93} \\
   LDA   &                 & 36  ~Ref. \onlinecite{ES99}  \\
   PBE   &                 & 8  ~Ref. \onlinecite{ES99}  \\
   SR-HF & 6-311G**/RI-J   & 16   ~Ref. \onlinecite{ZB_SR} \\
   SR-HF & DZVP/A2         & 16   ~Ref. \onlinecite{ZB_SR} \\
   SR-{\em Exact-Atomic}   & 6-311G**/RI-J & 19   ~Ref. \onlinecite{ZB_SR} \\
   SR-{\em Exact-Atomic}   & DZVP/A2       & 18   ~Ref. \onlinecite{ZB_SR} \\
   HFS (Uniform \af)       & 6-311G**/RI-J & 14  ~(Present work) \\
   HFS (Uniform \af)       & DZVP/A2       & 13  ~(Present work) \\
\hline
\end{tabular}
\end{table}

     To summarize, the performance appraisal of the analytic Hartree-Fock-Slater method 
is carried out for various \af values using the G2 database of 56 molecules.
The \af value that gives the best performance is determined by minimizing the 
mean absolute errors in the atomization energies of the G2 set of molecules. It
is shown that the analytic HFS model performs  better than the LDA or HF as well as the
SR method that uses atom dependent \af which give the exact HF or experimental
atomic energies. 
Further, by minimizing 
the MAE in the HF and the HFS total energies it is shown that the local exchange functional
performs significantly better for $\alpha = 0.7091 $ than the Gaspar-Kohn-Sham 
exchange functional. The MAE in former is an order of magnitude smaller than the MAE for 
the GKS exchange functional.  The use of this exchange functional in more sophisticated 
GGAs could  boost their performance considerably, and performance gain is already observed 
in case of Becke's exchange functional\cite{Handy01}. 

    Analytic DFT, even at this stage of development, is remarkably accurate.

        The Office of Naval Research, directly and through the Naval Research Laboratory, and and the 
Department of Defense's   High Performance Computing Modernization Program, through the Common High 
Performance Computing Software Support Initiative Project MBD-5, supported this work.


\begin{thebibliography}{99}

\bibitem{Cook1}
 K. S. Werpetinski and M. Cook,  Phys. Rev. A   {\bf 52}, 3397 (1995); 
 J. Chem. Phys.   {\bf 106}, 7124 (1997).

\bibitem{JF93} 
B. G. Johnson and M. J. Frisch, Chem. Phys. Lett. {\bf 216}, 133 (1993).


\bibitem{Pople98} 
J. A. Pople, Rev. Mod. Phys. {\bf 71}, 1267 (1998).

\bibitem{Kohn99}
W. Kohn, Rev. Mod. Phys. {\bf 71}, 1253 (1999).

\bibitem{Slater51} 
 J. C. Slater,  Phys. Rev. {\bf 81}, 385 (1951).

\bibitem{Dunlap03}  B. I. Dunlap, J. Phys. Chem. {\bf 107} 10082 (2003).

\bibitem{ZB_BN} R. R. Zope and B. I. Dunlap, Chem. Phys. Lett {\bf 386}, 403 (2004).

\bibitem{ZB_SR} R. R. Zope and B. I. Dunlap (unpublished). 


\bibitem{Dirac30}
 P. A. M. Dirac, Proc. Cambridge Philos. Soc. {\bf 26}, 376 (1930).

\bibitem{Gaspar} 
 G\'asp\'ar, R. Acta Phys. Hung.  {\bf 3}, 263 (1954). 

\bibitem{KS65} W. Kohn and L. J. Sham, Phys. Rev.  {\bf 140}  A1133 (1965).

\bibitem{Schwarz}
 K. Schwarz,  Phys. Rev. B {\bf 5}, 2466 (1972).

\bibitem{SS78}
 V. H. Smith and J. R. Sabin, J. Phys. B  {\bf 11}, 385 (1978).

\bibitem{Connolly}
 Connolly, J. W. D. In {\it  Modern Theoretical Chemistry;} Segal, G. A., Ed.; Plenum: New York, 1977;  Vol. 7, p 105.

\bibitem{Fliszar1}
 S. Flisz\'ar S.,  N. Desmarais, M.  Comeau,  J. Mol. Struct. (THEOCHEM)  {\bf 251}, 83 (1991);
 Vauthier, E. C., Coss\'e-Barbi, A.; Blain, M.; Flisz\'ar, S.
J. Mol. Struct. (THEOCHEM)  {\bf 492}, 113 (1999).

\bibitem{BR73}
 E. J. Baerends and P. Ros, Chem. Phys. {\bf 2}, 52 (1973).

\bibitem{BR78}
 E. J. Barends and P. Ros, Int. J. Quan. Chem. Symp. {\bf 12}, 169 (1978).

\bibitem{GHJ77}
 O. Gunnarson, J, Harris, and R. O. Jones, Phys. Rev. B  {\bf 15}, 3027 (1977).

\bibitem{DFT:Koch} W. Koch and M. C. Holthausen,
    {\it A Chemist's Guide to Density Functional Theory} (Wiley-VCH: Weinheim,
 Germany, 2001) p 74.

\bibitem{O1}
 R. Krishnan \etal,  J. Chem. Phys. {\bf 72}, 650, (1980).

\bibitem{O2}
 A. D. McLean, G. S. J. Chandler, J. Chem. Phys. {\bf 72}, 5639 (1980).

\bibitem{AW91}  J. Andzelm, E. Wimmer, J. Phys. B, 172, 307 (1991); J. Chem. Phys. {\bf 96}, 1280  (1992).

\bibitem{GSAW92}  N. Godbout \etal,  Can. J. Chem. {\bf 70}, 560  (1992).

\bibitem{Dunlap79} B. I. Dunlap, J. W. D. Connolly, and J. R. Sabin, J. Chem. Phys. {\bf 71}, 3396; 4993 (1979).

\bibitem{EWTR97}  K. Eichkorn \etal,  Theor. Chem. Acc. {\bf 97} (1997) 119.

\bibitem{BFGS} W. H. Press,  B. P. Flannery, S. A. Teukolsky, and W. T. Vetterling, 
    {\it Numerical Recipes The Art of Scientific Computing} (Cambridge University Press: Cambridge,
 England, 1986) p 309.

\bibitem{Pople1} 
J. A. Pople \etal,
 J. Chem. Phys. {\bf 90} 5662 (1989); 
  L. A. Curtiss \etal,
 J. Chem. Phys. {\bf 94} 7221 (1991).

\bibitem{Becke93} A. D. Becke, J. Chem. Phys.  {\bf 98} 1372  (1993).

\bibitem{Pople_DFT} B. G. Johnson, P. M. W. Gill, and J. A. Pople,
 J. Chem. Phys. {\bf 98} 5612 (1993).

\bibitem{ES99} M. Ernzerhof and G. E. Scuseria,
 J. Chem. Phys. {\bf 110} 5029 (1999).


\bibitem{Handy01}
N. C. Handy and A. J. Cohen,  Mol. Phys.  {\bf 99}, 403 (2001).

\end{thebibliography}
\end{document}